\documentclass[%
 reprint,
prl,
 amsmath,amssymb,
 aps,
]{revtex4-2}
\usepackage{overpic}
\usepackage{amsmath}
\usepackage{float}
\usepackage{graphicx}
\usepackage{subcaption}
\usepackage{dcolumn}
\usepackage{url}
\usepackage{xcolor}
\usepackage{hyperref}

\usepackage{hyperref}
\usepackage{appendix}
\usepackage{amsthm}
\usepackage[ruled,linesnumbered]{algorithm2e}
\hypersetup{
    colorlinks=true,  
    linkcolor=blue,   
    filecolor=blue,   
    urlcolor=blue,    
    pdftitle={Correlated error can impro},  
    pdfauthor={Author}, 
}
\begin{document}

\preprint{APS/123-QED}
\title{An exact Error Threshold of Surface Code under Correlated Nearest-Neighbor Errors: A Statistical Mechanical Analysis}

\author{SiYing Wang}
\author{ZhiXin Xia}
\author{Yue Yan}
\author{Xiang-Bin Wang}
\email{ xbwang@mail.tsinghua.edu.cn}
\affiliation{ State Key Laboratory of Low Dimensional Quantum Physics, Department of Physics, \\ Tsinghua University, Beijing 100084, China}

\date{\today}

\begin{abstract}
The surface code represents a promising candidate for fault-tolerant quantum computation due to its high error threshold and experimental accessibility with nearest-neighbor interactions. However, current exact surface code threshold analyses are based on the assumption of independent and identically distributed (i.i.d.) errors. Though there are numerical studieds for threshold with correlated error, they are only the lower bond ranther than exact value, this offers potential for higher error thresholds.Here, we establish an error-edge map, which allows for the mapping of quantum error correction to a square-octagonal random bond Ising model. We then present the exact threshold under a realistic noise model that combines independent single-qubit errors with correlated errors between nearest-neighbor data qubits. Our method is applicable for any ratio of nearest-neighbor correlated errors to i.i.d. errors. We investigate the error correction threshold of surface codes and we present analytical constraints giving exact value of error threshold. This means that our error threshold is both upper bound and achievable and hence on the one hand the existing numerical threshold values can all be improved to our threshold value, on the other hand, our threshold value is highest achievable value in principle.
\end{abstract}

\maketitle


\section{\label{sec:intro}introduction}
Quantum error correction codes are crucial for realizing quantum computation with noise\cite{quantum_codes1,quantum_codes2,quantum_codes3,quantum_codes4,quantum_codes5}. Surface code is a leading candidate due to two key advantages: it enables fault-tolerant computation using only nearest-neighbor gates in 2D, and it has a high error threshold\cite{Surface_code1,Surface_code_threshold1,Surface_code_threshold3,Surface_code_threshold4,xzzx_2021}. These properties make surface codes the basis for many quantum computer architectures across different platforms.\cite{experiment_correlated_2021,experiment_mid-circuit_2022,experiment_quantum_2022,experiment_google,experiment_quantum_ai_suppressing_2023,experiment_realization_2022,experiment_realizing_2022,experiment1,experiment2}.
\par
The threshold of the surface code reflects its maximum fault-tolerant capability and serves as a key technical indicator for quantum computing. Many studies have simulated surface code thresholds under realistic noise using decoding algorithms such as maximum likelihood decoder or minimum-weight perfect matching (MWPM)\cite{numerical_tomita_low-distance_2014,numerical_tomita_low-distance_2014,numerical_tiurev_correcting_2023,decoder_mps,decoder1,decoder2,decoder_important}. Though numerical studies exist for thresholds with correlated error\cite{numerical_fowler_quantifying_2014,numerical_nickerson_analysing_2019}, they provide only lower bounds rather than exact values, suggesting that higher error thresholds may be achievable.
\par
Knowing the precise threshold values plays a fundamentally important role in quantum error correction. Thresholds under correlated errors are particularly interesting because realistic quantum devices inevitably experience spatially and temporally correlated noise sources, such as qubit crosstalk during parallel operations\cite{correlated_crosstalk,correlated_crosstalk2}, leakage propagation between qubits\cite{quantum_codes1}, and non-Markovian environmental\cite{correlatedthreshold_breakdown_2014,correlatedthreshold_impact_2021}. These correlations can significantly deviate from the commonly assumed independent and identically distributed (i.i.d.) error models used in theoretical analyses\cite{correlatedthreshold_breakdown_2014,correlatedthreshold-chubb_statistical_2021,wang2025symmetry}.\par
A well known major problem is to work out the exact threshold values for surface codes under inter-qubit correlated errors. While i.i.d error models have well-established thresholds through connections to two-dimensional (2D) random-bond Ising model (RBIM)\cite{Surface_code1,surface_code_nishimori,Surface_code_threshold1,Surface_code_threshold3,Surface_code_threshold4,statistical_venn_coherent-error_2023}, the threshold result for inter-qubit correlated errors remains missing.
\par

In this work, we present exact threshold of surface codes under inter-qubit correlated errors. In particular, we study an inter-qubit correlated error model that combines nearest-neighbor correlations between data qubits with i.i.d errors on individual qubits. To manage the challenge arises from the difficulty of directly computing the probability of error chains, we first propose the error-edge mapping (EEM) method to model the probability of error chains. Using our EEM method, we can map the probability of successful error correction to partition function of a square-octagonal random bond Ising model. Give this fact, the problem of seeking the error threshold is transformed to calculate the critical point of the statistical mechanical model. 
\par
The paper is organized as follows: In Sec.\ref{sec:surface_code}, we review the error correction of surface codes and their relationship to critical phase transitions in statistical mechanics. In Sec.\ref{sec:error mode} we introduce the correlated error model under consideration. In Sec.\ref{sec:map}, we demonstrate how to map the surface code error correction problem with correlated errors to the phase transition problem of a statistical mechanical model. In Sec.\ref{sec:numerical}, we numerically calculate an example with correlated errors using Monte Carlo methods. Finally, we conclude the paper with a summary of our results.
\section{Error Correction in Surface Codes\label{sec:surface_code}}
The surface code is a topological quantum error correcting code that encodes logical qubits in a two-dimensional array of physical qubits arranged on a square lattice\cite{Surface_code1,surface_code_intro,Surface_code2}. Data qubits are placed on the edges, while ancilla qubits are located at vertices and plaquettes to measure stabilizer generators. The stabilizer group $\mathcal{S}$ is generated by X-stabilizers $A_v = \prod_{i \in \mathcal{N}(v)} X_i$ and Z-stabilizers $B_p = \prod_{j \in \mathcal{N}(p)} Z_j$, where $v$ denotes a vertex, $p$ denotes a plaquette, and $\mathcal{N}(\cdot)$ represents the set of neighboring data qubits. The logical operators are defined as $\bar{X} = \prod_{i \in \Gamma_X} X_i$ and $\bar{Z} = \prod_{j \in \Gamma_Z} Z_j$, where $\Gamma_X$ and $\Gamma_Z$ are chains spanning opposite boundaries.

Let $\mathcal{G}$ denote the stabilizer group of the surface code, which is generated by the X-stabilizers $A_v$ and Z-stabilizers $B_p$, i.e., $\mathcal{G} = \langle A_v, B_p \rangle$. The centralizer $\mathcal{C}(\mathcal{G})$ is defined as the set of all Pauli operators that commute with every element in $\mathcal{G}$, i.e., $\mathcal{C}(\mathcal{G}) = \{P : [P, g] = 0, \forall g \in \mathcal{G}\}$. This centralizer includes all stabilizer elements, logical operators, and their products.

When a Surface code experiences an error $E$, this error fails to commute with specific stabilizer generators $g$, which gives rise to a detectable syndrome pattern. The syndrome $S$ can be mathematically expressed as a binary vector where each component satisfies $\{s_i\in S:\forall g_i\in\mathcal{G}, E^{\dagger}g_iE=s_ig_i\}$.

The goal of quantum error correction is to determine a recovery operation $E’$ that generates an equivalent syndrome that produced by the original error $E$. Crucially, exact correspondence between $E’$ and $E$ is not required---any recovery operation satisfying $E'E\in \mathcal{C}(\mathcal{G})$ will exhibit syndrome equivalence. 

The success or failure of error correction depends on $E'E$. When $E'E$ belongs to $\mathcal{C}(\mathcal{G}) \backslash \mathcal{G}$, the decoding procedure introduces a logical error, indicating correction failure. In contrast, successful error correction occurs precisely when $E'E \in \mathcal{G}$.
\begin{figure}
    \centering
    \includegraphics[width=1\linewidth]{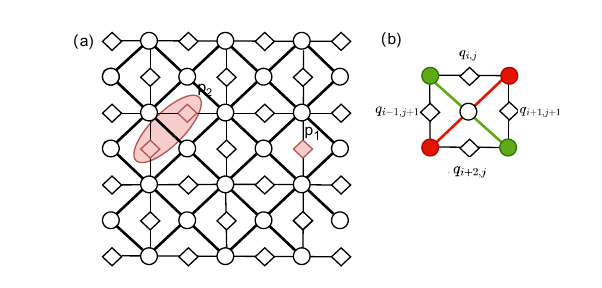}
    \caption{Error model. (a) $p_1$ represents the probability of a single qubit $Z$ error due to the channel in Eq.(\ref{eq:iid_error_model}) , while $p_2$ represents the probability of $ZZ$ error (enclosed by the pink ellipse) due to the channel in Eq.(\ref{eq:correlated_error_model}). In this diagram, neighboring ancilla qubits are connected by edges named in $l^k_2$ and $l^k_3$, defined in the main text letter around Eq.(\ref{eq:nE2}). For example, there are two edges of $l^k_2$ (in red) and $l^k_3$ (in green). (b)Four qubits are are denoted as $q_{i,j},q_{i-1,j+1},q_{i+1,j+1},q_{i+2,j}$ where the subscripts represent the spatial coordinates of each qubit's position, and they have be used in the main text around Eq.(\ref{eq:correlated_error_model}).}
    \label{fig:error_model}
\end{figure}
\section{Correlated errors between data qubits\label{sec:error mode}}
\begin{figure*}
    \centering
    \includegraphics[width=1\linewidth]{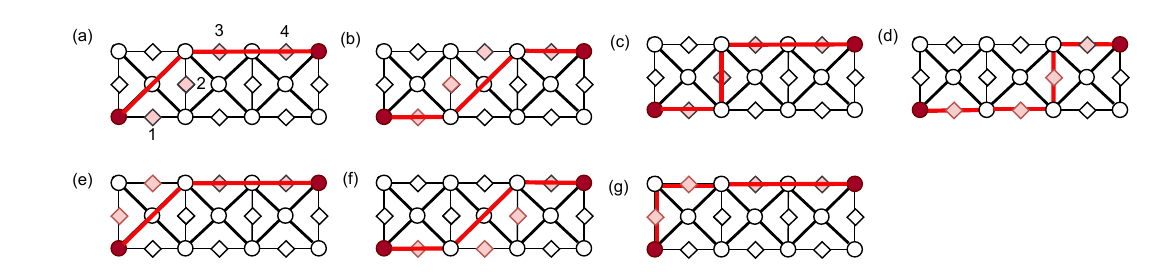}
    \caption{An example of error-edge map. Four pink diamonds indicate Z errors on the corresponding qubits. The figure shows 3 error sets containing different qubits, corresponding to 5 error chains. The error chains (red polylines) Fig.\ref{fig:error_case}(a) and (e) are the same, and we use $E'_1$ to indicate them. Consequently, the errors (pink diamonds) depicted in Fig.\ref{fig:error_case}(a) and Fig.\ref{fig:error_case}(e), they belong to the set $A_{1}$ according to our definitions for $E'_l,A_l$. Although the errors in Fig. 2(a), (b), and (c) occur on identical qubits, they belong to distinct equivalence classes $A_l$.}
    \label{fig:error_case}
\end{figure*}
We consider a noise model that combines i.i.d errors with correlated errors between nearest-neighbor data qubits. Since $X$ errors and $Z$ errors are corrected independently in surface codes, we focus our analysis on scenarios with only $Z$ errors. The error model are described as follows:
\begin{itemize}
    \item i.i.d phase flip error: Each data qubit independently experiences a Pauli error with probability $p_1$. \begin{equation}\label{eq:iid_error_model}
        \mathcal{E}(\rho)=\left(1-p_1\right) \rho+p_1 Z\rho Z^{\dagger}
    \end{equation}
    \item Correlated errors between nearest-neighbor data qubits: Each nearest data qubit pairs $\{ q_{i,j},q_{i-1,j+1}\}$ and $\{ q_{i,j},q_{i+1,j+1}\}$ has a probability $p_2$ of experiencing phase flip errors, which can be described as:
    \begin{equation}\label{eq:correlated_error_model}
        \mathcal{E}(\rho)=\left(1-p_2\right) \rho+p_2 Z_iZ_j\rho Z^{\dagger}_j Z^{\dagger}_i
    \end{equation}
\end{itemize}
In our model, both types of errors can occur simultaneously, and the resulting error patterns may overlap. The error model are depicted in Fig.(\ref{fig:error_model}).
\section{Mapping Correlated Errors to a Statistical model\label{sec:map}}
{The success probabilities of quantum error correction can be expressed as partition functions of statistical mechanical models, establishing a direct mapping between error thresholds and classical phase transitions\cite{Surface_code_threshold1,Surface_code1}}. First, the success probabilities of quantum error correction are:
\begin{align}\label{eq:p_success}
p_{success}&=\sum_{E'E \in \mathcal{G}} P(E'|S)=\frac{\sum_{E'E \in \mathcal{G}} P(E')}{\sum_{E'E\in \mathcal{C}(\mathcal{G})} P(E')}
\end{align}

where $E$ is a set of errors acting on the qubits and $P(E)$ is its corresponding probability of occurrence. However, our model contains complex inter-qubit correlated error, it is difficult to work out the probability $p(E)$ because $P(E)$ is a summation over numerous terms. For example, the total probability of generating the errors $E_1$ in Fig.\ref{fig:error_case}(a),(b),(c) is:
\begin{equation}
    P(E_1)\propto \frac{2p_2(1-p_2)}{1-2p_2(1-p_2)}\left(\frac{p_1}{1-p_1}\right)^2+\left(\frac{p_1}{1-p_1}\right)^4
\end{equation}
Below we introduce the \textit{error-edge map} developed in this work to simplify $P(E)$.\par
\textbf{Error-edge map: }
With the example shown in Fig.~\ref{fig:error_model}, we connect the nearest neighboring ancilla qubits of the surface code by thick black lines. We use $l^k_2$ to denote a diagonal edge (from upper-left to lower-right), and $l^k_3$ to denote an anti-diagonal edge (from upper-right to lower-left). Here the lowerscripts $2,3$ represent the edge types and the superscript $k$ represents the edge index. When correlated Z errors occur on either qubit pair $\{q_{i,j},q_{i-1,j+1}\}$ or qubit pair $\{q_{i+1,j+1},q_{i+2,j}\}$ (each with probability $p_2$), $n_{E'}(l^k_2) = 1$. When correlated errors happen on both pairs simultaneously (with probability $p_2^2$), they form a stabilizer operator $Z_{i+1,j+1}Z_{i+2,j}Z_{i,j}Z_{i+1,j+1}$, and hence we treat this case as having no error. Similarly, when correlated Z errors occur on either qubit pair $\{q_{i,j},q_{i+1,j+1}\}$ or $\{q_{i-1,j+1},q_{i+2,j}\}$, the green edge $l^k_3$ is marked with $n_{E'}(l^k_3) = 1$; when correlated errors occur on both pairs, we regard this case as having no error. \par
\begin{figure}
    \centering
    \includegraphics[width=0.8\linewidth]{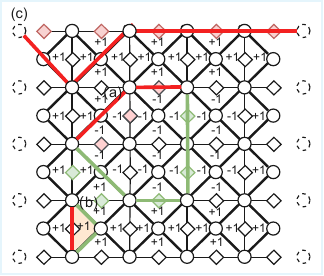}
    \caption{Example of error correction on a surface code. Pink diamonds indicate Z errors occurring on the corresponding qubits, while green diamonds represent Pauli Z operators for recovery of the respective qubits. Red lines represent the error chain $E$, and green lines represent the recovery chain $E'$. The dashed circles represent virtual ancilla qubits. After performing our error-edge map, connected error chains corresponding to connections between two virtual nodes (dashed dots) will produce line $L$ that span the surface code. At the centers of triangulars, Ising spins take values in $\{\pm 1\}$. (a) $E + E' = C$ represents a perfect error correction example, where $C$ encloses a domain with spin value $-1$. (b) An example of ising spins corresponding to unwanted terms obtaining in partition function mapped by $\mathcal{M}$. (c) A failed error correction example where $E+ E' = L$ and L span the surface code.}
    \label{fig:cycle}
\end{figure}

Before presenting the simplified expression of $P(E)$ by our error-edge map, we first introduce some mathematical notations. We use $l_1^{k}$ to denote a horizontal edge or a vertical edge,  $n_E(l_1^{k})$ to indicate whether an error occurs on the data qubit at edge $l_1^{k}$, then we have:
\begin{align}\label{eq:nE2}
\displaystyle n_E( l_i^{k}) =\begin{cases}
0 & l_i^{k}\notin \tilde{E}\\
1 & l_i^{k}\in \tilde{E}
\end{cases}
\end{align}
where $i\in\{1,2,3\}$ and $\tilde{E}$ denotes the set of edges containing qubits that experience errors. The notation $n_E( l_1^{k})$ will be used later in Eq.(\ref{PE}). With these notations we are ready to write the simplified expression of $P(E)$ by our error-edge map. Applying the rule in the paragraph above to all qubits surrounding each plaquette,  we obtain the probability of recovery chain $\tilde{E}'$ can be written as:
\begin{equation}\label{PE}
    P(\tilde{E}') \propto \prod_{i\in\{1,2,3\}}\prod _{l^k_{i}}\left(\frac{\bar{p}_{i}}{1-\bar{p}_i}\right)^{n_{E'}( l^k_{i})}
\end{equation}
where $\bar{p}_1=p_1,\bar{p}_{2}=2(1-p_2)p_2,\bar{p}_{3}=2(1-p_3)p_3$, $\tilde{E}'$ denotes the set of edges that satisfy $n_{E'}(l^k_{i}) = 1$. Next, we will prove that
\begin{align}\label{eq:sum_PE}
    &\sum_{\tilde{E}' + \tilde{E} \in \{C\}} p(\tilde{E}') = \sum_{E'E \in \mathcal{G}} p(E')\\
    &\sum_{\tilde{E}' + \tilde{E} \in \{L\}} p(\tilde{E}') = \sum_{E'E \in \mathcal{C}(\mathcal{G}) \backslash \mathcal{G}} p(E').   
\end{align}
Here $C$ represents a closed cycle, and $L$ denotes a line traversing the surface code, as shown in Fig.\ref{fig:cycle}.\par
\textit{Proof: }Suppose that the syndrome $S$ is fixed, $E_l$ is a set of errors that lead to the syndrome $S$. Let $Z_u$ be the Z errors occurring on the qubit or qubit pair corresponding to $u$, where $u$ is the index of a qubit or a qubit pair that experiences errors. For example, as shown in Fig.\ref{fig:error_case}(a), $E=\{Z_u|u \in \{(1,2), 3, 4\}\}$, $Z_{(1,2)}$ represents that qubit pair $(1,2)$ experiences $Z$ error with probability $p_2$, and $Z_3,Z_4$ represents that qubits 3 and 4 experience $Z$ error with probability $p_1$. We define $A_{l}=\{\mathcal{E}_l,E_l\}$, where
\begin{equation}
    \mathcal{E}_l=\{\mathcal{E}^{(i,j)}_l| \forall (i,j) \text{ s.t. }Z_{(i,j)} \in E_l\}
\end{equation}
and $\mathcal{E}^{(i,j)}_l$ is
\begin{equation}
   \{Z_u|Z_u \in E_l\setminus\{Z_{(i,j)}\}\} \cup \{Z_u|\exists B_p \text{ s.t. } B_pZ_{(i,j)}\in E_l\}
\end{equation}
where $Z_{(i,j)}$ is the correlated error in $E_l$ induced by Eq. (\ref{eq:correlated_error_model}). Taking Fig.\ref{fig:error_case} as an illustrative example, we can readily observe that $A_l$ contains errors that share the same error chain $\tilde{E}'$ (as indicated by the red polylines in Fig.\ref{fig:error_case}). An example is given in the caption of Fig.\ref{fig:error_case}. 
\par
After taking the error-edge map as stated earlier, we obtain $P(\tilde{E}'|S)=\sum_{E' \in A_{l}} P(E'|S)$. Note that elements in $A_l$ can differ from each other only by a stabilizer. Therefore, given any two errors $E_1, E_2 \in A_{l}$, it follows that they must be in the same coset of $\mathcal{G}$ which is defined in Sec.\ref{sec:surface_code}. Therefore
\begin{align}
    \sum_{\tilde{E}' + \tilde{E}\in \{C\}}p(\tilde{E}'|S) = \sum_{\mathcal{X}}\sum_{E \in A_{l}} p(E|S)
\end{align}
where $\mathcal{X}=A_{l}| A_{l}E\in\mathcal{G}$, $A_{l}E$ represents the new set obtained by multiplying each element in set $A_{l}$ with $E$. This completes the proof of Eq. (\ref{eq:sum_PE}).

\begin{figure}
    \centering
    \includegraphics[width=1\linewidth]{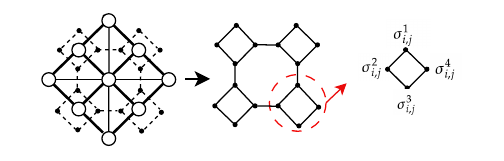}
    \caption{Mapping a surface code to a square-octagonal lattice.}
    \label{fig:map_lattice}
\end{figure}

To carry out the mapping that follows, we introduce $\eta$ and $u$:
\begin{align}\label{eq:eta_E}
\displaystyle \eta_E( l^k_{i}) &=\begin{cases}
1 & l^k_{i}\notin \tilde{E}\\
-1 & l^k_{i}\in \tilde{E}
\end{cases}\\
\displaystyle u(l^k_{i}) &=\begin{cases}
1 & l^k_{i}\notin \tilde{E}'+\tilde{E}\\
-1 & l^k_{i}\in \tilde{E}'+\tilde{E}
\end{cases}
\end{align}
Substituting $n_{E'}(l^k_{i})$ to $\eta_{E'}( l^k_{i})$, we have
\begin{align}\label{eq:pE1_EXP}
    p(\tilde{E}') &\propto \prod_{i\in\{1,2,3\}} \prod_{l^k_{i}} \left(\frac{1-\bar{p}_i}{\bar{p}_i}\right)^{\frac{1}{2}\eta_{E'}(l^k_{i})} \nonumber \\
    &= \exp\left(\sum_{i\in\{1,2,3\}} \sum_{l^k_{i}} \frac{1}{2}J_i\eta_{E'}(l^k_{i})\right)
\end{align}
Here $J_i=\ln\left(\frac{1-\bar{p}_i}{\bar{p}_i}\right)$. Observe that:
\begin{equation}
        \eta_{E'}(l^k_{i})=u(l^k_{i})\eta_E(l^k_{i})
\end{equation}
Then Eq.\ref{eq:pE1_EXP} can be written as
\begin{equation}
        p(\tilde{E}') \propto \exp\left(\sum_{i\in\{1,2,3\}} \sum_{l^k_{i}} \frac{1}{2}J_iu(l^k_{i})\eta_{E}(l^k_{i})\right)
\end{equation}
Consider placing a lattice point $\sigma_{i,j}^k \in \{\pm 1\}$ at the center of each triangular in Fig.\ref{fig:cycle}. This gives us the dual lattice structure illustrated in Fig.\ref{fig:map_lattice}. With this we can define our map $\mathcal{M}$ through Eq.\ref{eq:map}:
\begin{align}\label{eq:map}
    \mathcal{M}: \begin{cases}
    u(l^k_1) \mapsto \begin{cases}
\sigma_{i,j}^4\sigma^1_{i+1,j} & \text{if } l^k_1 \text{ is a horizontal edge} \\
\sigma_{i,j}^3\sigma^2_{i,j+1} & \text{if } l^k_1 \text{ is a vertical edge}
\end{cases}\\
    u(l^k_2)\mapsto-\frac{1}{2}(\sigma_{i,j}^1\sigma_{i,j}^2-1)(\sigma_{i,j}^3\sigma_{i,j}^4-1)+1\\
    u(l^k_3)\mapsto-\frac{1}{2}(\sigma_{i,j}^1\sigma_{i,j}^3-1)(\sigma_{i,j}^2\sigma_{i,j}^4-1)+1
    \end{cases}
\end{align}
\begin{figure*}
    \centering
    \includegraphics[width=1\linewidth]{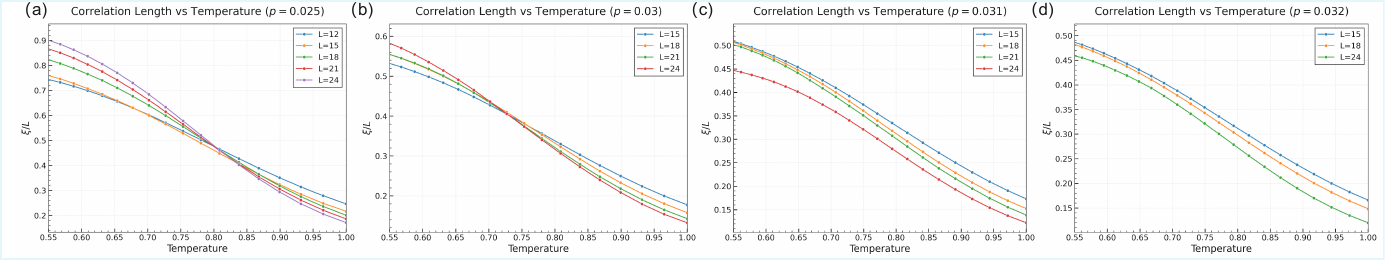}
    \caption{Finite-size correlation length $\xi_m/L$ as a function of temperature $T$ for different error probabilities. The temperature range is $T \in [0.5, 1]$. (a): $p = 0.025$. (b): $p = 0.030$. (c): $p = 0.031$. (d): $p=0.032$. For $p \lesssim p_c$ there is clear evidence of a phase transition (curves for different system sizes $L$ cross) whereas for $p > p_c$ the transition disappears.}
    \label{fig:threshold_simulation}
\end{figure*}
Given map $\mathcal{M}$, the cycle $\tilde{E} +\tilde{E}' = C$ corresponds to domain walls on the dual lattice, as shown in Fig.\ref{fig:cycle}. After mapping, the partition function contains unwanted terms, as shown in Fig.\ref{fig:cycle}(b). Therefore, we impose constraints by Eq.(\ref{eq:constraint}) to eliminate the unwanted terms so as to ensure the mapped partition function yields correct results:
\begin{equation}\label{eq:constraint}
    \sigma_{i,j}^1\sigma_{i,j}^2\sigma_{i,j}^3\sigma_{i,j}^4=1
\end{equation}
Applying map $\mathcal{M}$ (Eq.\eqref{eq:map}) to Eq.\ref{eq:pE1_EXP} with the constraint of Eq.\ref{eq:constraint}, we have
\begin{align}\label{eq:Ising model}
     p(E') \propto Z[&J_i,\eta_{E}(l^k_{i})]\\\nonumber
     =\sum_{\sigma_{i,j}^k}\exp\bigg\{&\frac{1}{2}J_1\eta_{E}(l^k_1)(\sigma_{i,j}^3\sigma_{i+1,j}^1+\sigma_{i,j}^4\sigma_{i,j+1}^2)\\\nonumber
     +&\frac{1}{2}J_2\eta_{E}(l^k_2)(\sigma_{i,j}^1\sigma_{i,j}^2+\sigma_{i,j}^3\sigma_{i,j}^4)\\\nonumber
     +&\frac{1}{2}J_3\eta_{E}(l^k_3)(\sigma_{i,j}^1\sigma_{i,j}^4+\sigma_{i,j}^2\sigma_{i,j}^3)\bigg\}
\end{align}
Eq.\ref{eq:Ising model} describes the partition function of square-octagonal random bond ising model. Parameter $\eta_{E}(l^k_{i})$ determines whether these interactions are attractive or repulsive.\par
Computing the exact threshold of the surface code can be transformed into calculating the phase transition point given by Hamiltonian $H$. According to Eq. (\Ref{eq:Ising model}) and (\ref{eq:constraint}), we obtain:
\begin{align}
    H = &-\sum_{i,j}\bigg[\eta_{E}(l^k_1)(\sigma_{i,j}^3\sigma_{i+1,j}^1+\sigma_{i,j}^4\sigma_{i,j+1}^2)\\\nonumber
     &+J'_2\eta_{E}(l^k_2)(\sigma_{i,j}^1\sigma_{i,j}^2+\sigma_{i,j}^3\sigma_{i,j}^4)\\\nonumber
     &+J'_3\eta_{E}(l^k_3)(\sigma_{i,j}^1\sigma_{i,j}^4+\sigma_{i,j}^2\sigma_{i,j}^3)\bigg]\\\nonumber
     & \text{with constraints: }\sigma_{i,j}^1\sigma_{i,j}^2\sigma_{i,j}^3\sigma_{i,j}^4=1
\end{align}
here $J'_2=J_2/J_1, J'_3=J_3/J_1$. When $p_2$ approaches 0, both $J_2$ and $J_3$ tend to positive infinity, which implies that $\sigma_{i,j}^1,\sigma_{i,j}^2,\sigma_{i,j}^3,\sigma_{i,j}^4$ must have the same value, either $+1$ or $-1$, and hence the statistical model degenerates to the random bond Ising model with i.i.d. disorder as shown in \cite{Surface_code1}.

The free energy cost of non-trivial domain walls is
\begin{equation}\label{eq:free_energy}
\Delta_i(\tau) = \beta F(K, \tau_i) - \beta F(K, \tau) = \ln \left(\frac{Z[K, \tau]}{Z[K, \tau_i]}\right)
\end{equation}
where $\beta = 1/T$ is the inverse temperature, $\beta=J_1=-\frac{1}{2}\ln(\frac{p_1}{1-p_1})$, which corresponds to the Nishimori line. Here $K$ represents the coupling strength $J'_2,J'_3$ apearing in Eq.\ref{eq:Ising model}, and $\tau_i$ denotes the configuration with a domain wall inserted compared to the reference configuration $\tau$.
The transition free energy cost in Eq.\ref{eq:free_energy} diverges as $L \to \infty$ in the ordered phase, while it converges to a finite constant in the disordered phase.

If the error rate is below threshold,  the system is in the ordered phase. Forming a non-trivial domain wall requires infinite free energy according to Eq.\ref{eq:free_energy}, which prevents logical operators from penetrating through the surface code and hence corresponds to successful error correction. If the error rate is above threshold, thermal fluctuations dominate, resulting in a disordered phase, which means non-trivial domain walls can be created without energy cost, leading to error correction failure.
\section{Calculating thresholds with Monte Carlo methods\label{sec:numerical}}
Here we consider the case where both error rates present in Sec.\ref{sec:error mode} are equal: $p_{2} = p_{1} = p$.
We perform parallel tempering Monte Carlo simulations across various system sizes and temperature ranges\cite{Monte_earl2005parallel,Monte_predescu2005efficiency,Montecarlo1}, with the parameters detailed in Table \ref{para}.
The critical temperature is determined through finite-size scaling analysis of the correlation length. We define the wave-vector-dependent magnetic susceptibility as \cite{Surface_code1,Surface_code_threshold1,threshold_colorcode_error_2009}:
\begin{figure}\label{fig:Stim}
    \centering
    \includegraphics[width=1\linewidth]{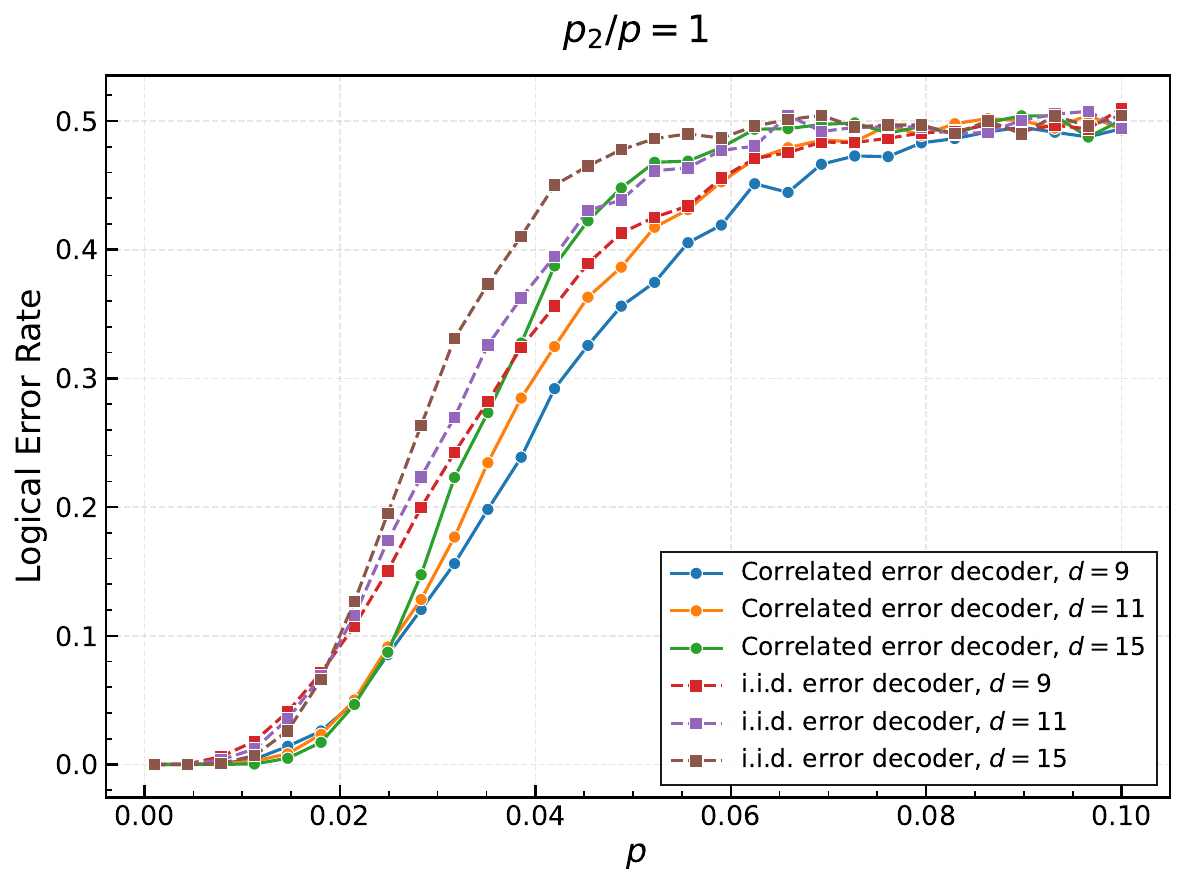}
    \caption{Logical error rate of the $d=9,15,21$ surface codes as a function of error probability $p$. The dashed lines (square points) represent the decoding results when error correlations are ignored and correlated errors are simply treated as i.i.d. errors, with a threshold of approximately 1.9\%. The solid lines (circular points) represent the decoding results using a decoder specifically designed for correlated errors, with a threshold of approximately 2.4\%.}
\end{figure}
$$\chi(\vec{k}) = \frac{1}{L^2} \left\langle \left| \sum_i s_i e^{i\vec{k} \cdot \vec{r}_i} \right|^2 \right\rangle$$

where $L$ is the linear system size, $s_i$ are the Ising spins, $\vec{r}_i$ denotes the spatial position of site $i$, and $\langle \cdot \rangle$ represents thermal and disorder averaging.

The finite-size correlation length is then computed by:

$$\xi = \frac{1}{2\sin(k_{min}/2)} \sqrt{\frac{\chi(\vec{0})}{\chi(\vec{k}_{min})} - 1}$$

where $\vec{k}_{min} = (2\pi/L, 0)$ is the minimal non-zero wave vector.

Near the phase transition, the normalized correlation length exhibits finite-size scaling below:
\begin{equation}\label{eq:scaling}
    \frac{\xi}{L} \approx f\left[L^{1/\nu}(T - T_c)\right]
\end{equation}

where $f$ is a dimensionless scaling function, $\nu$ is the correlation length critical exponent, and $T_c$ is the critical temperature. At the critical temperature $T_c$, $\xi/L$ in Eq.\eqref{eq:scaling} becomes size-independent, causing curves for different system sizes to intersect, as shown in Fig. \ref{fig:threshold_simulation}(a). Absence of such intersections indicates no phase transition in the studied temperature range.
\begin{table}
\centering
\caption{Simulation parameters. $L$ : linear system size. $N_{\text{sa}}$ : the number of disorder samples, $t_{\text{eq}} = 2^b$: maximum number of equilibration sweeps. The simulation is stopped earlier if our equilibrium criterion is met., $T_{\min}$ ($T_{\max}$) : the lowest (highest) temperature, $N_T$: the number of temperatures used. We use $L = \{12, 15, 18, 21, 24\}$.}
\label{para}
\begin{tabular}{ccccccc}
\hline\hline
$p$ & $L$ & $N_{\text{sa}}$ & $b$ & $T_{\min}$ & $T_{\max}$ & $N_T$ \\
\hline
0.025 & 12,15,18 & 600 & 21 & 0.3 & 1 & 30 \\
0.025 & 21,24 & 600 & 21 & 0.5 & 1 & 20 \\
0.035 & 15,18 & 600 & 21 & 0.3 & 1 & 30 \\
0.035 & 21,24 & 600 & 21 & 0.5 & 1 & 20 \\
0.045 & 15,18 & 600 & 21 & 0.3 & 1 & 35 \\
0.045 & 21,24 & 600 & 21 & 0.5 & 1 & 20 \\
0.030--0.034 & 15,18 & 800 & 21 & 0.5 & 1& 25 \\
0.030--0.034 & 21,24 & 220 & 21 & 0.5 & 1& 25 \\ \\
\hline\hline
\end{tabular}
\end{table}
To determine if the system has reached equilibrium, we use logarithmic binning of the data and perform an equilibrium test every 10,000 sweeps\cite{eq_test1,eq_test2}. The simulation is terminated early if two consecutive tests are passed. Following this, we perform 200,000 measurement sweeps, collecting data every 5 sweeps. The results are shown in Fig.\ref{fig:threshold_simulation}. The calculated threshold is 3\%.

To compare our calculated exact threshold with existing decoding methods, we take Pymatching 2.0\cite{pymatching2,higgott2022pymatching}. In running Pymatching 2.0, we take $p_1=p_2$, as calculated in Monte Carlo results by our own method. We take two ways in applying pymatching decoding. We first ignore the correlated errors in the system and directly use a decoder designed for i.i.d. noise in decoding, the threshold is 1.8\%. We then use Stim's \textit{detector error model} to generate minimum weight matching graphs under correlated errors for decoding, the threshold is 2.4\%.  This remains 0.5\% lower than our calculated threshold 3\%, indicating significant room for improvement in decoding correlated errors.

\section{Conclusion\label{sec:conclusion}}
In this work, we have investigated the error correction threshold of surface codes under a realistic noise model that combines independent single-qubit errors with spatially correlated errors between nearest-neighbor data qubits. By employing the statistical mechanical mapping approach, we mapped the quantum error correction problem onto a square-octagonal random bond Ising model, enabling exact threshold calculations  independent of specific decoding algorithms.

Our numerical simulation shows that existing decoders still exhibit a significant gap from our exact threshold in the presence of correlated errors. This indicates plenty of room for existing decoders to be further improved when correlated errors exist.

\begin{acknowledgements}
We acknowledge the financial support in part by National Natural Science Foundation of China grant No.12174215 and No.12374473, and Innovation Program for Quantum Science and Technology No.2021ZD0300705. This study is also supported by the Taishan Scholars Program. 
\end{acknowledgements}
\bibliographystyle{unsrt}
\bibliography{ref.bib}
\end{document}